\documentclass[a4paper,11pt]{article}
\pdfoutput=1 

\usepackage{jheppub} 
\usepackage{amsfonts}
\usepackage{amssymb,amsmath}
\usepackage{hyperref}
\usepackage{comment}
\usepackage{graphicx}
\usepackage{subcaption}
\usepackage{color,graphicx,slashed,multirow,booktabs,multirow}
\usepackage{lipsum}
\usepackage{braket}
\usepackage{bbold}
\usepackage{xcolor}
\usepackage[utf8]{inputenc}
\usepackage{subfloat}
\usepackage{cleveref}
\usepackage{ulem}
\usepackage{eqparbox}
\newcommand{\be}{\begin{equation}}
\newcommand{\ee}{\end{equation}}
\newcommand{\bea}{\begin{eqnarray}}
\newcommand{\eea}{\end{eqnarray}}
\newcommand{\ba}{\begin{aligned}}
\newcommand{\ea}{\end{aligned}}
\usepackage{mathtools}

\usepackage{orcidlink} 

\usepackage[force]{feynmp-auto}
\usepackage[T1]{fontenc} 

\newcommand{\equaref}[1]{Eq.~(\ref{#1})}

\newcommand{\secref}[1]{Section~\ref{#1}}

\title{Monopoles at Future Neutrino Detectors}
\preprint{IPPP/25/20}

\author[a]{Pablo M. Candela~\orcidlink{0009-0009-8416-9295}\,,}
\affiliation[a]{Instituto de F\'{i}sica Corpuscular (CSIC-Universitat de Val\`{e}ncia), Parc Cient\'ific UV C/ Catedr\'atico Jos\'e Beltr\'an, 2 E-46980 Paterna (Valencia) - Spain}

\author[b]{Valentin V. Khoze~\orcidlink{0000-0003-3006-4147}\,,}
\affiliation[b]{Institute for Particle Physics Phenomenology, Department of Physics, Durham University, Durham DH1 3LE, U.K.}

\author[b]{Jessica Turner~\orcidlink{0000-0002-9679-5252}}

\emailAdd{pamuca@ific.uv.es}
\emailAdd{valya.khoze@durham.ac.uk}
\emailAdd{jessica.turner@durham.ac.uk}

\abstract{
We investigate the potential of future neutrino experiments, DUNE and Hyper-Kamiokande, to probe magnetic monopoles via Callan-Rubakov (CR) processes. We consider both relativistic and non-relativistic monopoles and focus on two primary detection signatures: high-energy antiproton production and proton decay catalysis. For relativistic monopoles, our analysis of the CR process indicates antiproton production with energies near 900~GeV and we find that both experiments can provide limits on the fluxes an order of magnitude below the Parker bound (approximately \(\Phi \lesssim  10^{-16}\,\mathrm{cm^{-2}\,s^{-1}\,sr^{-1}}\)). For non-relativistic monopoles, we recast the experimental sensitivity to proton decay catalysis and obtain upper limits on the monopole flux of \(\Phi \lesssim 2.3 \times 10^{-23}\,\mathrm{cm^{-2}\,s^{-1}\,sr^{-1}}\) for Hyper-Kamiokande and \(\Phi \lesssim 1.1 \times 10^{-22}\,\mathrm{cm^{-2}\,s^{-1}\,sr^{-1}}\) for DUNE.
 }


\begin{document}

\maketitle

\section{Introduction}
\label{sec:introduction}
Magnetic monopoles were first proposed to explain the quantisation of electric charge \cite{Dirac:1931kp}; however, they also naturally emerge as topological defects formed during the spontaneous symmetry breaking of Grand Unified Theories (GUTs) in the early Universe \cite{tHooft:1974kcl, Polyakov:1974ek}. Consequently, the possible discovery of monopoles would enhance our understanding of cosmology and fundamental symmetries \cite{Zeldovich:1978wj,Preskill:1979zi,Guth:1980zm}, which has motivated various searches over the last few decades. Experimental strategies to detect magnetic monopoles are broadly categorised as direct or indirect. Direct searches aim to detect monopoles produced in high-energy collisions at particle accelerators \cite{MoEDAL:2019ort, Gould:2024zed}, or to identify their characteristic ionisation or Cherenkov radiation as they traverse detectors \cite{IceCube:2021eye, NOvA:2020qpg, IceCube:2014xnp}. Indirect searches rely on astrophysical phenomena, notably monopole-induced nucleon decay, where the decay products, such as neutrinos, provide observational signatures \cite{Super-Kamiokande:2012tld, Zhang:2024mze}. A general constraint on the monopole abundance, independent of their velocity, is provided by the Parker bound \cite{Parker:1970xv}. This constraint comes from the requirement that monopoles must not significantly drain energy from galactic magnetic fields over long timescales, providing an upper monopole flux limit of \( \Phi \lesssim 10^{-15} \, \text{cm}^{-2} \, \text{s}^{-1} \, \text{sr}^{-1} \). This limit corresponds to one monopole traversing a square-metre area every three hundred years. Velocity-dependent constraints complement this more general limit by targeting monopoles in specific kinematic regimes. 
The IceCube experiment provides stringent constraints on relativistic monopoles with velocities \(0.75 < \beta < 0.995\), giving flux limits as restrictive as \( \Phi \lesssim 2.0 \times 10^{-19} \, \text{cm}^{-2} \, \text{s}^{-1} \, \text{sr}^{-1} \) for monopole masses \( m_M \sim 10^{8-11} \, \text{GeV} \) \cite{IceCube:2021eye}. 

The MACRO experiment was the first to search for GUT-scale monopoles, searching for energy deposition from monopoles with velocities \(4 \times 10^{-5} < \beta < 1\). MACRO set a flux limit of \( \Phi_{90} \leq 1.4 \times 10^{-16} \, \text{cm}^{-2} \, \text{s}^{-1}\) at 90\% confidence level (C.L.).  IceCube also searched for non-relativistic monopole constraints by investigating monopole-catalysed proton decay through processes such as \(M + p \rightarrow M + e^+ + \pi^0\), achieving an upper bound on the monopole flux of \( \Phi_{90} \leq 10^{-18} \, (10^{-17}) \, \text{cm}^{-2} \, \text{s}^{-1} \, \text{sr}^{-1}\) for catalysis cross-sections of \(10^{-22} \, (10^{-24}) \, \text{cm}^2\) \cite{IceCube:2014xnp}. More recently, experiments such as NOvA have provided constraints for monopoles with velocities \(6 \times 10^{-4} < \beta < 5 \times 10^{-3}\) and masses above \(m_M > 5 \times 10^8\,\text{GeV}\), setting flux limits of \(\Phi_{90} < 2 \times 10^{-14} \, \text{cm}^{-2} \, \text{s}^{-1} \, \text{sr}^{-1}\) \cite{NOvA:2020qpg}. Additionally, recent theoretical analyses leveraging constraints such as the Andromeda Parker bound have further tightened flux limits, achieving \( \Phi \lesssim 5.3 \times 10^{-19} \, \text{cm}^{-2} \, \text{s}^{-1} \, \text{sr}^{-1} \) for monopole masses \( m_M \sim 10^{13-16} \, \text{GeV} \) \cite{Zhang:2024mze}. Finally, Super-Kamiokande has provided exceptional sensitivity by searching for neutrino signals from proton decays catalysed by non-relativistic monopoles trapped in celestial bodies. The particularly sensitive channel \(p \to \mu^+ K^0\) generates monoenergetic muon neutrinos ($\sim100$ MeV), allowing Super-Kamiokande to achieve a flux limit of \( \Phi \lesssim 10^{-24}\, \text{cm}^{-2} \, \text{s}^{-1} \, \text{sr}^{-1}\) for monopoles with \(\beta \sim 10^{-3}\) \cite{Hu:2022wcd}.

In this work, we first investigate Callan-Rubakov (CR) processes induced by relativistic and non-relativistic magnetic monopoles at future neutrino experiments, specifically DUNE \cite{DUNE:2020ypp} and Hyper-Kamiokande \cite{Hyper-Kamiokande:2018ofw}. We focus on the CR process that generates high-energy antiprotons with energies around 900 GeV. We demonstrate that the effective detection volume of each experiment can be significantly enhanced by approximately $238\%$ for DUNE and $25\%$ for Hyper-Kamiokande, substantially increasing the expected event rate. Notably, Hyper-Kamiokande can uniquely detect both the high-energy antiprotons produced by CR processes and the Cherenkov radiation signatures from relativistic monopoles. Assuming maximal CR cross-sections, we find that the sensitivity of Hyper-Kamiokande and DUNE to monopole flux falls roughly an order of magnitude below the Parker bound after 15 years of operation.

Next, we recast the projected sensitivity of the proton decay catalysed by non-relativistic monopoles in both experiments. Our analysis shows that for monopoles with velocities around $\beta=10^{-3}$, the absence of observed proton decay events would translate into stringent flux upper limits of approximately $1.1 \times 10^{-22}~\mathrm{cm^{-2}\,s^{-1}\,sr^{-1}}$ and $1.5 \times 10^{-23}~\mathrm{cm^{-2}\,s^{-1}\,sr^{-1}}$ for DUNE and Hyper-Kamiokande, respectively. We also highlight that, unlike GUT-mediated proton decay, which produces back-to-back decay products, in monopole-induced proton decay, the decay products would be forward-going, providing an experimental method to differentiate these processes. 

This paper is structured as follows: In \secref{sec:monopole-theory}, we discuss monopole interactions via Callan-Rubakov processes for both antiproton synthesis (for relativistic monopoles) and proton catalysis (for non-relativistic monopoles). We present detailed kinematics for antiproton synthesis, clarifying what regions of the monopole mass and energy parameter space future neutrino detectors would be sensitive to, and we investigate these effects for various monopole charges $q_J = \lbrace1/2,\, 3/2,\, 6/2\rbrace$. Additionally, we discuss monopole energy loss during passage through the Earth and its implications for the monopole energies at the detectors. In \secref{sec:data-analysis}, we estimate relevant backgrounds (atmospheric neutrino-induced muons), detail our signal event calculations for high-energy antiprotons, and describe the method to effectively expand the detector volume to include the surrounding material. Subsequently, we perform a background-free analysis, presenting our results and flux upper limits for relativistic and non-relativistic monopoles in \secref{sec:results-and-discussion}. We summarise and make concluding remarks in \secref{sec:conclusion}.

\section{Monopole interactions with fermions via Callan-Rubakov processes}
\label{sec:monopole-theory}
It is well-known since the early eighties that magnetic monopoles, $M$, can catalyse baryon plus lepton ($B+L$) number violation in the Standard Model (SM) by scattering on SM fermions~\cite{Rubakov:1981rg,Rubakov:1982fp,Callan:1982ah,Callan:1982au}. 
Two types of such $(B+L)$-violating Callan-Rubakov type processes are particularly relevant to us. The first is monopole-induced proton decay,
\begin{equation}
\label{eq:ptoe}
M\,+\, p\,\rightarrow\, M\,+\, e^+\,+\,X\,,
\end{equation}
and the second is the inverse process: antiproton synthesis,
\begin{equation}
\label{eq:etop}
M\,+\, e\,\rightarrow\, M\,+\, \bar{p}\,+\,X\,.
\end{equation}
Both such processes change $B+L$ by two units. These are inclusive processes in the sense that the specified $(B+L)$-violating two-particle final states are accompanied by an additional radiation $X$ which preserves all perturbative symmetries of the system, \textit{i.e.}, has vanishing electric charge, baryon and lepton numbers. In the single-family approximation, $X$ represents pions ($\pi^0$ and $\pi^+ \pi^-$ pairs) and, when other families are included $X$ is
expected to include other mesons and leptons, particularly kaons and muons~\cite{Rubakov:1988aq}.
\subsection{Antiproton synthesis}
\label{sec:p_synth}
Scattering cross-sections for the reactions \equaref{eq:ptoe} and \equaref{eq:etop} were recently studied in Ref.~\cite{Khoze:2024hlb} and we will follow their analysis. First, consider the antiproton synthesis process at the $2\to2$ particle level, ignoring $X$ in the final state. The differential cross-section for this process is given by~\cite{Khoze:2024hlb}
\begin{equation}
\label{eq:sig_etop}
M\, e_R^- \rightarrow M\, \overline{p}_L \quad :\quad
  \dfrac{\mathrm{d}\sigma}{\mathrm{d}\Omega} \,=\, \dfrac{1}{2} \dfrac{|\mathbf{p_{\overline{p}}^{\, cm}}|}{|\mathbf{p_{e}^{\, cm}}|} \dfrac{q_J^2}{|\mathbf{p_{e}^{\, cm}}|^{2}} \left[\sin\left(\dfrac{\theta^{\mathrm{cm}}}{2}\right)\right]^{4 |q_J|-2},
\end{equation}
where the subscripts $R$ and $L$ indicate the electron and antiproton helicities, $\mathbf{p_{i}^{\, cm}}$ ($\theta^{\mathrm{cm}}$) denotes the three-momentum (scattering angle) in the centre-of-mass (CoM) frame of particle $i$ and $q_J$ is half of the monopole magnetic charge in units of $2 \pi \hbar /e$. It is worthwhile to comment on the various factors in this formula briefly. 

The Dirac quantisation condition implies that $q_J$ must take values $q_J \in \mathbb{Z} / 2$ so that the magnetic monopole $M_1$ with a minimal magnetic charge has $q_J=1/2$. In this paper, along with $M_1$, we will also consider magnetic monopoles $M'_3$ and $M'_6$ with $q_J=3/2$ and $q_J=3$, respectively.\footnote{All these are stable monopoles in the Standard Model and are consistent with the electroweak symmetry breaking. But unlike the minimal monopole $M_1$, the monopoles $M'_3$ and $M'_6$ carry no long-range chromomagnetic fields and, being colour-singlets, provide more suitable than $M_1$ asymptotic states that can hit the Earth from the outer space. We refer to~\cite{Khoze:2024hlb} for more detail on the monopole species with SM quantum numbers.} The cross-section scales as $q_J^2$, as it should, since it is the coupling constant between the electric and magnetic degrees of freedom. Furthermore, note that the cross-section is not suppressed by the monopole mass or by any Grand Unification scale -- this is the central feature of all Callan-Rubakov processes, which in the infrared are governed by the strong interactions scale $\sim 1\, {\rm GeV}$ rather than the much higher new physics scale, so that $\sigma \sim q_J^2 / ({\rm GeV}^2).$ We also note that the Callan-Rubakov processes occur only in the lowest available partial wave given by $j_0= |q_J|-1/2$ in the angular momentum decomposition; all higher partial wave scatterings do not change $B+L$ nor flip chiralities of the (massless) SM fermions between the in and out states.

The cross-section formula in~\equaref{eq:sig_etop}
is presented in the centre-of-mass frame with $\mathbf{p_{e}^{\, cm}}$, $\mathbf{p_{\overline{p}}^{\, cm}}$ respectively denoting the incoming electron and the outgoing antiproton three-momenta, and $\theta^{\mathrm{cm}}$ is the scattering angle in this frame. The process is kinematically allowed only above the energy threshold $|\mathbf{p_{\overline{p}}^{\, cm}}| \ge 0$ for the final-state antiproton to be on-shell; below that value the cross-section vanishes. The factor $0\le |\mathbf{p_{\overline{p}}^{\, cm}}|/(2|\mathbf{p_{e}^{\, cm}}|)<1/2$ in~\equaref{eq:sig_etop} can be viewed as the branching ratio for the final state $M\, \overline{p}_L $ discriminating between the two underlying minimal processes,
\begin{equation}
\label{eq:sig_etwo}
M\, e_R^- \rightarrow M\, \overline{p}_L \qquad {\rm and}\qquad
M\, e_R^- \rightarrow M\, {e}_L^-\,.
\end{equation}
The latter process exists already in QED and was computed by Kazama, Yang and Goldhaber (KYG) in the seventies~\cite{Kazama:1976fm} by solving the non-relativistic quantum mechanical scattering of an electron on an infinitely heavy magnetic monopole.
Stripping off the branching ratio, the cross-section in~\equaref{eq:sig_etop} agrees with the KYG result,
$q_J^2 / |\mathbf{p_{e}}|^{2} \left[\sin(\theta/2)\right]^{4 |q_J|-2}$,
for the second process in~\equaref{eq:sig_etwo}.
This expression for the cross-section, including its dependence on the scattering angle in the CoM frame, is also in agreement with the relativistic generalisation of the KYG scattering that was derived in~\cite{Csaki:2020inw} using the pairwise helicity formalism.\footnote{See also Refs.~\cite{Csaki:2021ozp,Khoze:2023kiu,vanBeest:2023dbu,vanBeest:2023mbs} for more detail on scattering amplitudes of monopoles on fermions.}

To produce the antiproton in the final state, the electron momentum in the CoM frame should be above the kinematic threshold,
$|\mathbf{p_{e}^{\, cm}}| > m_p$ (for the realistic case of a heavy monopole $m_M \gg m_p$). If we allow for additional mesons $X$ in the final state, \textit{e.g.} $\pi^+ \pi^-$ pairs, the kinematic threshold will be pushed up by the relevant meson masses, $2m_\pi$ for each pair.
The cross-sections have a power-law decrease in the high energy regime and hence receive dominant contributions at energies not much above the kinematic threshold ({\it cf.}~top row panels in Fig.~\ref{fig:cs-and-momenta-comparison}). It follows that the most favourable final states would involve the lowest thresholds, and one can justifiably ignore contributions of $X$ when estimating rates for inclusive processes~\equaref{eq:etop}. With this in mind, we will use the cross-section formula in~\equaref{eq:sig_etop} for the antiproton synthesis processes inside neutrino detectors catalysed by an incoming relativistic monopole.

In \secref{sec:kin-synth}, we will examine the monopole kinematics for the antiproton synthesis process and point out that to produce an on-shell antiproton; the monopole has to be highly relativistic, with the Lorentz gamma factor at least $\sim 2 \times 10^3$.
Monopoles produced after the phase transition\footnote{Phase transitions relevant for monopole production are expected to occur in theories where the hypercharge \( U(1)_Y \) group of the Standard Model is embedded in a non-Abelian group within a UV-complete theory. This embedding may or may not be part of a Grand Unified Theory with a single simple gauge group.} in the early Universe
can be rapidly accelerated by galactic and extra-galactic magnetic fields extended over large spatial domains~\cite{Wick:2000yc,Perri:2023ncd}, hence acquiring kinetic energies of the order of  $E^{\rm kin}_M \sim 10^{14} \,{\rm GeV}$. This implies that monopoles with masses \( m_M \lesssim 10^{14} \,{\rm GeV} \) become relativistic and can trigger antiproton synthesis. Conversely, heavier GUT-scale monopoles would remain non-relativistic and could be responsible for catalysed proton decay.

\subsection{Catalysed proton decay}
\label{sec:proton-decay}
A non-relativistic monopole can be very efficient in inducing a rapid proton decay in the collision with a stationary proton
via~\equaref{eq:ptoe}. These are classic Callan-Rubakov processes with the
cross-section given by (see~\cite{Rubakov:1988aq} for a classic review and~\cite{Khoze:2024hlb} for a recent discussion of the cross-section estimate),
\begin{equation}
\label{eq:sig_pdec}
M\, p_L \rightarrow M\, \overline{e}_R +X \quad :\quad
  \dfrac{\mathrm{d}\sigma}{\mathrm{d}\Omega} \,=\, \dfrac{q_J^2}{\beta^2} \,\dfrac{1}{m_p^2} \left[\sin\left(\dfrac{\theta^{\mathrm{cm}}}{2}\right)\right]^{4 |q_J|-2}\,.
\end{equation}
This estimate holds for non-relativistic monopoles with velocity $\beta = v/c < 0.2$ in the laboratory frame where the proton is at rest.
The cross-section is enhanced at small values of $\beta$ with an unphysical singularity at $\beta \to 0$, which should be cut off in the IR at 
$\beta_{IR} \sim 10^{-5} - 10^{-3}$.

\subsection{Kinematics of the monopole scattering for the antiproton synthesis}
\label{sec:kin-synth}
Consider the process $M\, e^- \rightarrow M\, \overline{p}$ in which a monopole $M$ scatters off an electron $e^-$ in the lowest partial wave, resulting in the production of an antiproton $\overline{p}$ with the resulting change in the baryon plus lepton numbers by two units, $\Delta (B+L)=2$. The kinematic threshold for this reaction occurs when the kinematic invariant
$s$ reaches its minimal value allowed by kinematics, $s_{\mathrm{th}}=(m_M+m_p)^2$, and corresponds to both final-state particles being at rest in the CoM frame. The physical frame for the experimental setup is the laboratory frame where a relativistic extragalactic monopole hits a (nearly) stationary orbital electron, which is taken to be at rest. In this frame, the kinematic invariant $s$ is 
\begin{equation}
\label{eq:s-lab}
s\,=\, \left(E^{\mathrm{lab}}_M +m_e\right)^2 - \left(\mathbf{p_{M}^{\, lab}}\right)^2\,=\, 2m_e E^{\mathrm{lab}}_M +m_e^2+m_M^2\,,
\end{equation}
and for the monopole energy, we have
\begin{equation}
\label{eq:EM-lab}
E^{\mathrm{lab}}_{M} \,=\, \dfrac{1}{2m_e} \left(s-m_M^2-m_e^2\right) \,\ge\,
E_{\mathrm{th}}\,,
\end{equation}
where we have defined $E_{\mathrm{th}}$ as the threshold 
the energy of the monopole in the laboratory frame. This threshold energy is given by
\begin{equation}
\label{eq:Eth-lab}
  E_{\mathrm{th}} = \dfrac{2 m_p m_M + m_p^2 - m_e^2}{2 m_e} \approx \dfrac{m_p}{m_e} m_M \,\simeq\, 1.8 \times 10^3\, m_M\,.
\end{equation}
It is clear that in order to produce the antiproton in the final state $E_M^{\mathrm{lab}} > E_{\mathrm{th}}$, the monopole must be relativistic with the Lorentz factor
$\gamma \gtrsim 2000$.
\begin{figure}
  \centering
  \includegraphics[width=\textwidth]{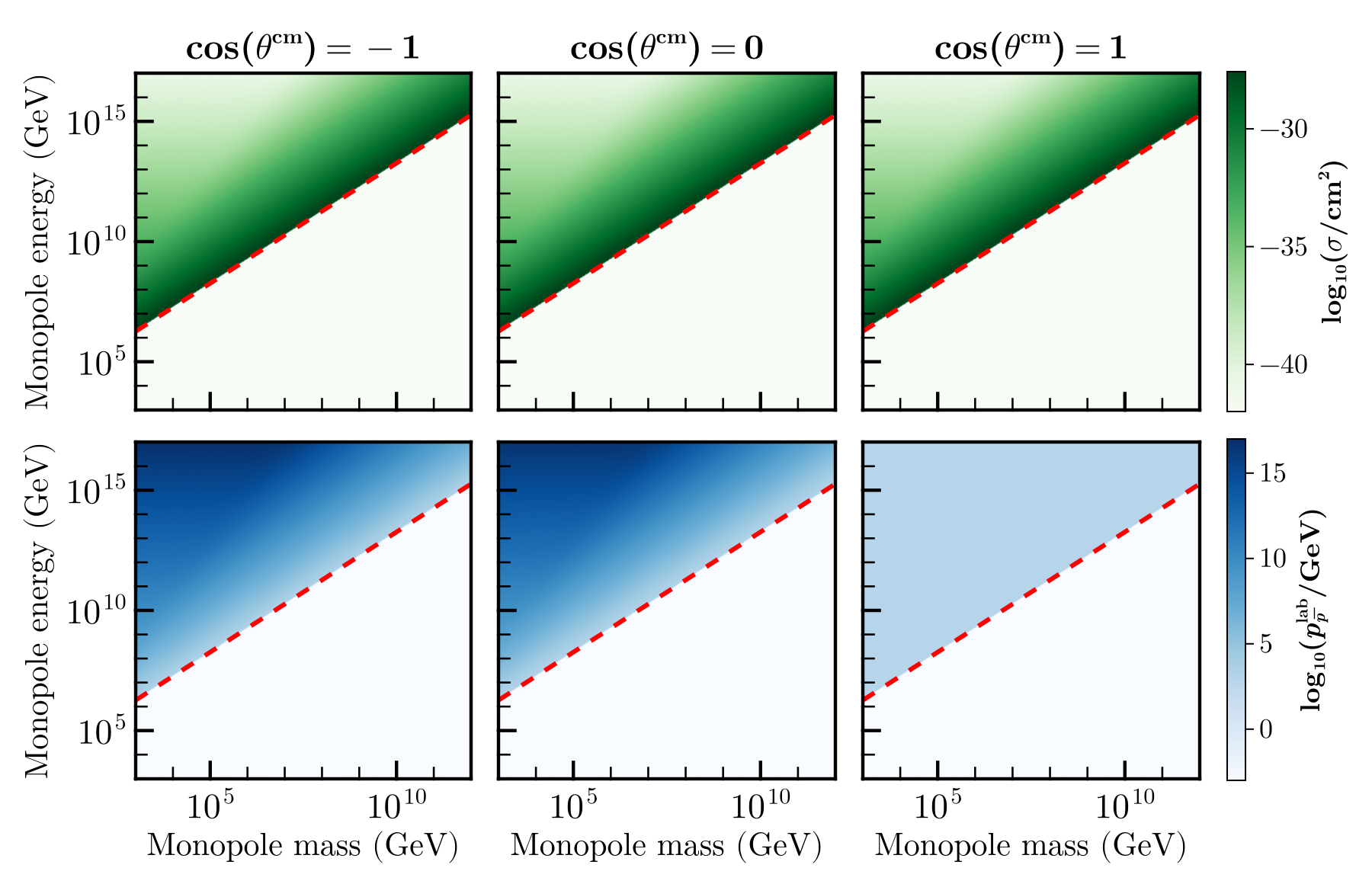}
  \caption{Top row: Logarithm (base 10) of the total cross-section (with $q_J = 1/2$) for different monopole energies (measured at the detector in the laboratory frame) and masses. Bottom row: Logarithm (base 10) of the final three-momentum of the antiproton in the laboratory frame for different monopole energies and masses. We have considered three scattering angles in the centre-of-mass frame, ranging from backward to forward scattering. The threshold energy limit in the laboratory frame is marked as a dashed red line.}
 \label{fig:cs-and-momenta-comparison}
\end{figure}
This produces some interesting effects on the antiproton in the laboratory frame, depending on the scattering angle in the centre of mass frame\footnote{We take this to be the angle between the incoming electron and the outgoing antiproton, or equivalently, the angle between the incoming and the outgoing monopole in the centre-of-mass frame.}, $\theta^{\mathrm{cm}}$. To better illustrate this, let us plot the magnitude of the three-momentum of the antiproton in the laboratory frame as a function of the monopole mass and energy. In the bottom row of Fig.~\ref{fig:cs-and-momenta-comparison}, we have represented the logarithm in base 10 of the final antiproton momentum for different monopole masses and energies. These are the monopole energies at the detector in the laboratory frame. When $\cos(\theta^{\mathrm{cm}}) = 1$, $\theta^{\mathrm{cm}} = 0$, and the monopole continues straight in the centre-of-mass frame after the interaction. If one boosts the system back to the laboratory frame, this scenario results in the monopole also moving straight after the collision. We refer to this configuration as forward scattering. In this case, the antiproton three-momentum is relatively small, approximately $10^3~\mathrm{GeV}$.

Conversely, when $\cos(\theta^{\mathrm{cm}}) \neq 0$, and we transform back to the laboratory frame, the monopole scatters at an angle after the collision. Since the monopole is a massive object moving at a high velocity and the antiproton is significantly lighter, altering the monopole's trajectory requires the antiproton to carry a substantial amount of momentum. This results in large antiproton momenta in cases where the monopole does not undergo forward scattering, as observed in the first two columns of the bottom row of Fig.~\ref{fig:cs-and-momenta-comparison}. For completeness, the case $\cos(\theta^{\mathrm{cm}}) = -1$ corresponds to backward scattering, wherein the monopole recoils and moves in the opposite direction post-scattering.
\begin{figure}
  \centering
  \begin{subfigure}{0.49\textwidth}
    \includegraphics[width=\textwidth]{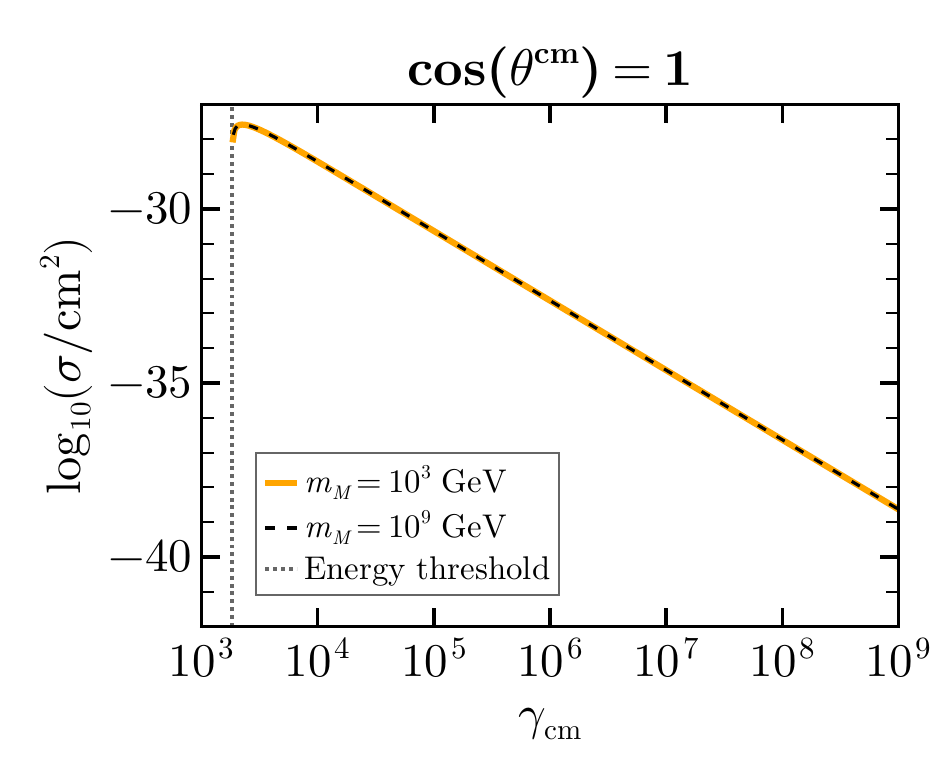}
  \end{subfigure}
  \hfill
  \begin{subfigure}{0.49\textwidth}
    \includegraphics[width=\textwidth]{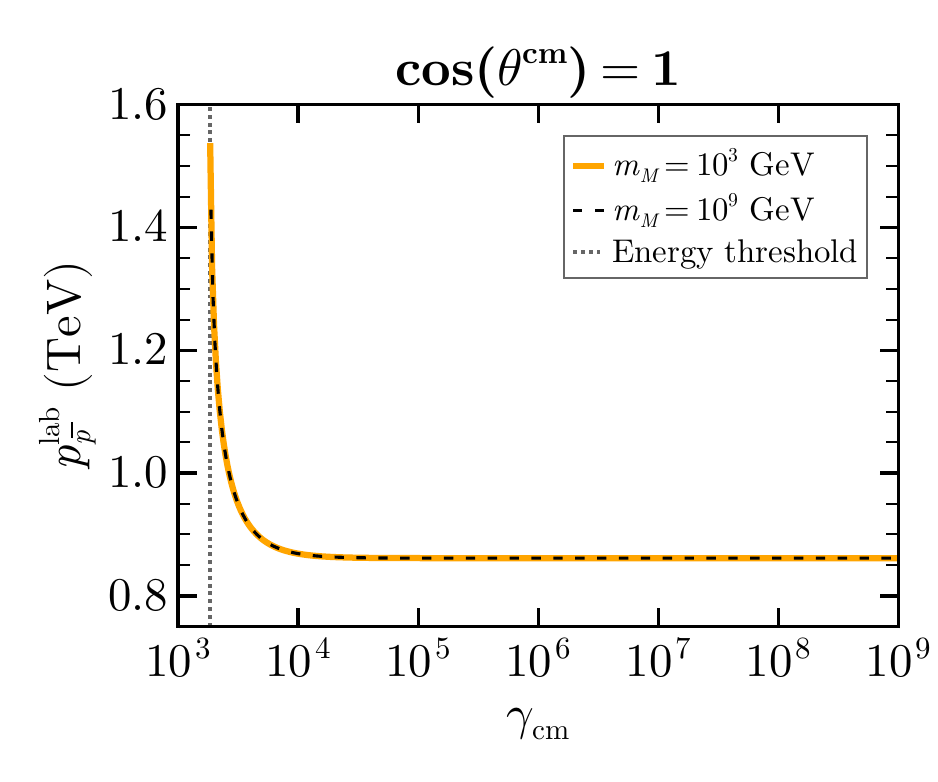}
  \end{subfigure}
  \caption{Left: Total cross-section (with $q_J = 1 / 2$) as a function of the Lorentz factor of the centre-of-mass frame for the forward scattering case. Right: Final antiproton momenta in the laboratory frame as a function of the Lorentz factor of the centre-of-mass frame for the forward scattering case. The energy threshold of the process for $m_M = 10^3~\mathrm{GeV}$ is indicated as a dotted gray line.}
  \label{fig:cs-vs-gamma}
\end{figure}
The top row of Fig~\ref{fig:cs-and-momenta-comparison} shows the logarithm in base 10 of the cross-section in units of $\rm{cm}^{2}$ for different monopole masses and energies for the case $q_J = 1 / 2$. In this instance, the cross-section of Eq.~\eqref{eq:sig_etop} is independent of the scattering angle, and therefore, the three subplots show the same pattern where the cross-section peaks near the threshold energy. The cross-section decreases as one moves away from the threshold energy, represented by a dashed red line almost perpendicular to it. Thus, let us further explore this behaviour.

Let us first define $\beta_{\mathrm{cm}}$ as the velocity needed to boost from the laboratory frame to the centre-of-mass frame in which the collision will occur. The Lorentz factor of the centre-of-mass frame is therefore $\gamma_{\mathrm{cm}} = 1 / \sqrt{1 - \beta_{\mathrm{cm}}^2}$. In the left panel of Fig.~\ref{fig:cs-vs-gamma} we have represented the total cross-section (with $q_J = 1/2$) of Eq.~\eqref{eq:sig_etop} against this Lorentz factor $\gamma_{\mathrm{cm}}$. We fix two different monopole masses of $10^3$ GeV and $10^9$ GeV and then vary their energies, which changes the gamma factor. The cross-sections peak close to the energy threshold, represented by the dotted grey line, and then decrease logarithmically as we increase the energy. The result is the same independently of the monopole mass chosen\footnote{This is in agreement with the general structure of CR processes, which dictate that the scattering rates are largely insensitive to the monopole mass and are unsuppressed by any BSM unification scales.}. 

This explains the darker band, almost parallel to the threshold energy, of the top row of Fig.~\ref{fig:cs-and-momenta-comparison} (dashed red line) and why the colours become much brighter as we move further away from it.

In the right panel of Fig.~\ref{fig:cs-vs-gamma}, we show the value of the momentum of the final antiproton in the laboratory frame versus the Lorentz factor of the centre-of-mass for two fixed monopole masses. Notably the result is again independent of the mass. In addition, we have found that at higher energies, the value of the momentum plateaus at around $861.4$ GeV. This is why the plot seems to have only one solid colour in the rightmost figure of the bottom row of Fig.~\ref{fig:cs-and-momenta-comparison}. Therefore, no matter the scattering angle or the energy of the monopole, as long as the process is kinematically allowed, the final state will produce a relativistic antiproton of at least $861.4$ GeV in the laboratory frame, which has a distinctive signal in neutrino experiments. Moreover, because magnetic monopoles carry electric charge, any relativistic monopole traversing the water volume of Hyper‑Kamiokande will emit Cherenkov radiation when its velocity exceeds the phase velocity of light in water \cite{PhysRev.138.B248}. This distinctive Cherenkov light signature, combined with the uniquely identifiable antiproton signal, provides a powerful, multi‑channel signal to discriminate monopole events from the background.

\subsection{Monopole energy loss outside of the detector}
\label{sec:outside-detector}
\begin{figure}[t!]
  \centering \includegraphics[width=0.45\linewidth]{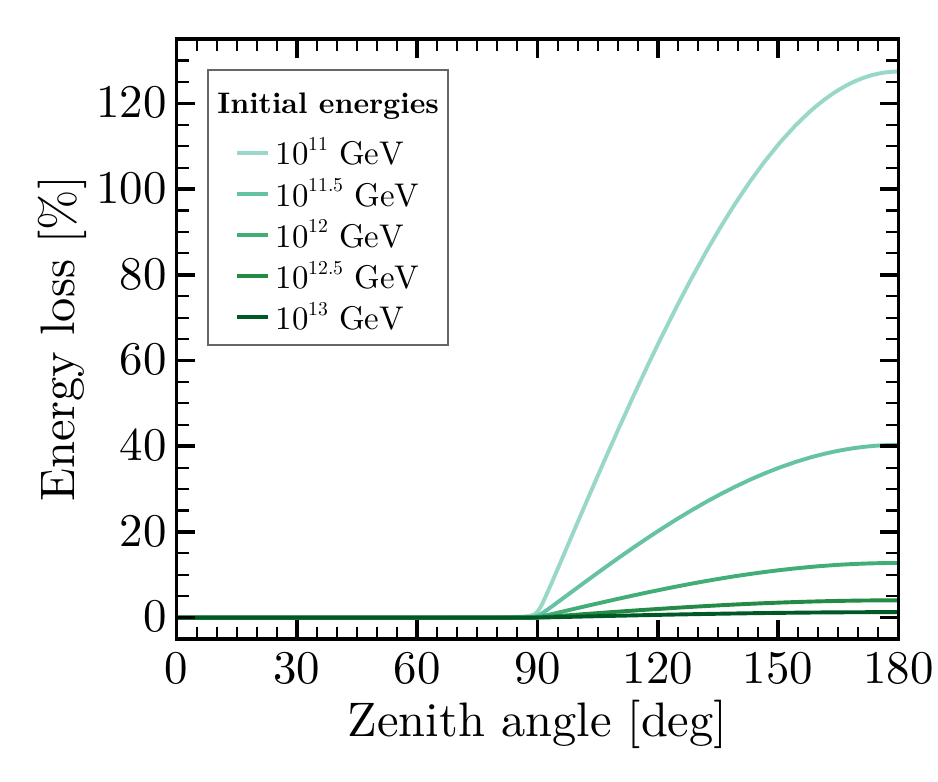}
\includegraphics[width=0.45\linewidth]{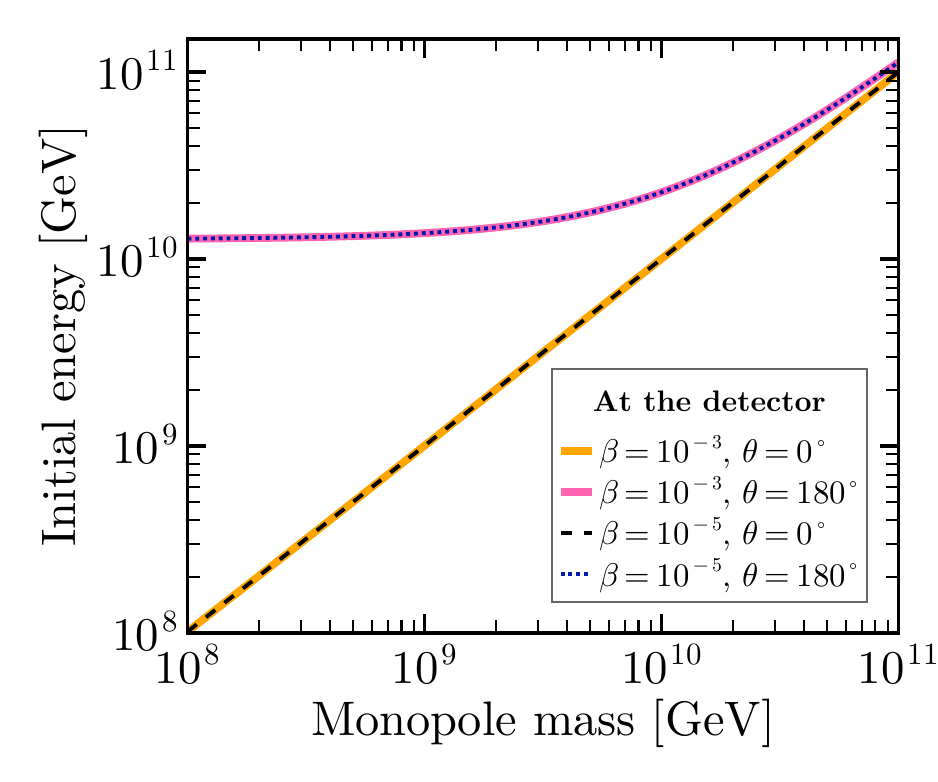}
  \caption{Left: Percentage energy loss as a function of the incident zenith angle for different initial energies of relativistic monopoles, once they arrive to the detector. Right: initial energies for different masses of non-relativistic monopoles, given a certain velocity $\beta$ and angle $\theta$ needed for them to arrive at the detector.}
  \label{fig:energyloss}
\end{figure}
Monopoles arriving at the neutrino detectors will have traversed different path lengths depending on their arrival direction: a downward-going monopole passes only through the detector’s overburden ($\sim1.5$ km in rock for DUNE and $\sim0.8$ km for Hyper‑Kamiokande), whereas an upward-going monopole must cross nearly the Earth’s full diameter ($\sim1.27\times10^4$~km). Because the magnetic charge is conserved and monopole masses far exceed those of nucleons or electrons, monopoles can neither decay nor be significantly deflected as they traverse the Earth. However, they lose energy in matter, which increases with the path length. The atmospheric energy losses ($\sim10^5\,\mathrm{GeV}$) of the monopole are negligible in comparison with the energy loss through the planet. The dominant energy loss arises from $(B+L)$‑preserving ionisation interactions with electrons of the Earth. Stopping powers in rock (and iron) lie between $10-100\,\mathrm{GeV\,cm^{-1}}$~\cite{Derkaoui:1998uv}, implying a maximum loss of $\sim10^{11}\,\mathrm{GeV}$ for a monopole crossing its diameter. 

In the left panel of Fig.~\ref{fig:energyloss}, we present the percentage energy loss of a monopole as it travels from the surface of the Earth to a detector located 1.5 km underground. We assume an energy loss of 100~GeV cm$^{-1}$ for various monopole energies at the Earth’s surface. The energy loss remains minimal (at the percent level) until the monopole comes from the horizon, \textit{i.e.}, with a zenith angle of $\theta = 90^\circ$. For example, with an energy of $E_M = 10^{11.5}$~GeV, an upward-going monopole (with $\theta = 180^\circ$) would have its energy be reduced roughly by a factor of 2. On the other hand, the percentage of energy loss is smaller for monopoles of even higher energy, even at larger zenith angles, which correspond to longer path lengths.

The minimum energy required for a downward-going monopole to reach the detector is approximately \(E_M \sim 10^8\) GeV, corresponding to an energy loss of roughly 15\% upon detection. In contrast, an upward-going monopole with the same initial energy would not reach the detector. For a monopole arriving from the opposite side of the Earth, an initial energy of approximately \(10^{10}\) GeV at the Earth's surface is insufficient to ensure detection. However, if the monopole's energy is increased to about \(E_M \sim 10^{11.2}\) GeV, it can be detected despite an energy loss of around 80\%. Monopoles are accelerated to energies of up to \(10^{14}\) GeV by extragalactic fields. Accordingly, the present analysis applies to monopole energies between \(10^{11}\) GeV and \(10^{14}\) GeV. This energy bound may be lowered to around \(10^8\) GeV if only downward- and horizontally incident monopoles are considered, as monopoles below this threshold will not reach the detector. Moreover, kinematic constraints require the monopole's energy to be at least about 2000 times its mass, restricting the sensitivity of this study to monopole masses in the range \(10^{5}\) GeV \(\lesssim m_M \lesssim 10^{11}\) GeV.

Finally, let us note that energy losses via Callan–Rubakov (CR) interactions, \textit{e.g.}, Eq.~\eqref{eq:etop}, is subdominant. With a mean free path of $\lambda \gtrsim 70~\mathrm{m}$ (see Appendix.~\ref{app:MFP}), a monopole traveling through the whole Earth undergoes $\sim 10^5$ CR collisions (versus $\sim 20$ CR collisions if it arrives from above), each removing $\sim 10^3~\mathrm{GeV}$. Thus, total CR losses ranges from $10^4\,\mathrm{GeV}$ to $10^8\,\mathrm{GeV}$, which is three orders of magnitude below ionisation losses\footnote{Equivalently, combining a number $n_{\mathrm{CR}}$ of CR collisions, $M + n_{\mathrm{CR}}\cdot e \to M + n_{\mathrm{CR}}\cdot \overline{p}$, in Eq.~\eqref{eq:Eth-lab} yields $\Delta E_M^{\mathrm{lab}}\approx n_{\rm CR}\cdot10^3\,\mathrm{GeV}$.}.
Throughout our analysis, we will consider the energy of the monopoles to be monochromatic \textit{at} the detector and assume an isotropic flux similar to Ref.~\cite{IceCube:2021eye}. This monopole energy range also applies to the non-relativistic case, with the difference that there is no energy threshold for the catalysed proton decay, as opposed to the antiproton production. We can estimate the typical masses and energies at the surface of the Earth that result in non-relativistic monopoles arriving at the detector. In the right panel of Fig.~\ref{fig:energyloss} we show how the initial energies of the non-relativistic monopoles must relate to their masses given an arrival velocity at the detector for a given zenith angle. We note that the energy loss for non-relativistic monopoles is lower than that of relativistic ones \cite{Derkaoui:1998uv}, and we take this to be $10$ GeV cm$^{-1}$. We observe that initial higher energies at the surface are required for monopoles traversing the entire diameter of the Earth. Moreover, as anticipated, higher initial monopole energies are required for higher masses.

\section{Analysis}
\label{sec:data-analysis}
As discussed at the end of \secref{sec:kin-synth}, a relativistic antiproton is produced after the monopole scatters off. If there are no other particles with such high energies or they have a signal that can be easily rejected as background, compared to the antiproton production, one can perform a simple analysis assuming that the process is background-free.
\subsection{Background estimation}
\label{sec:background-estimation}
\begin{figure}[t!]
  \centering
  \begin{subfigure}{0.49\textwidth}
    \includegraphics[width=\textwidth]{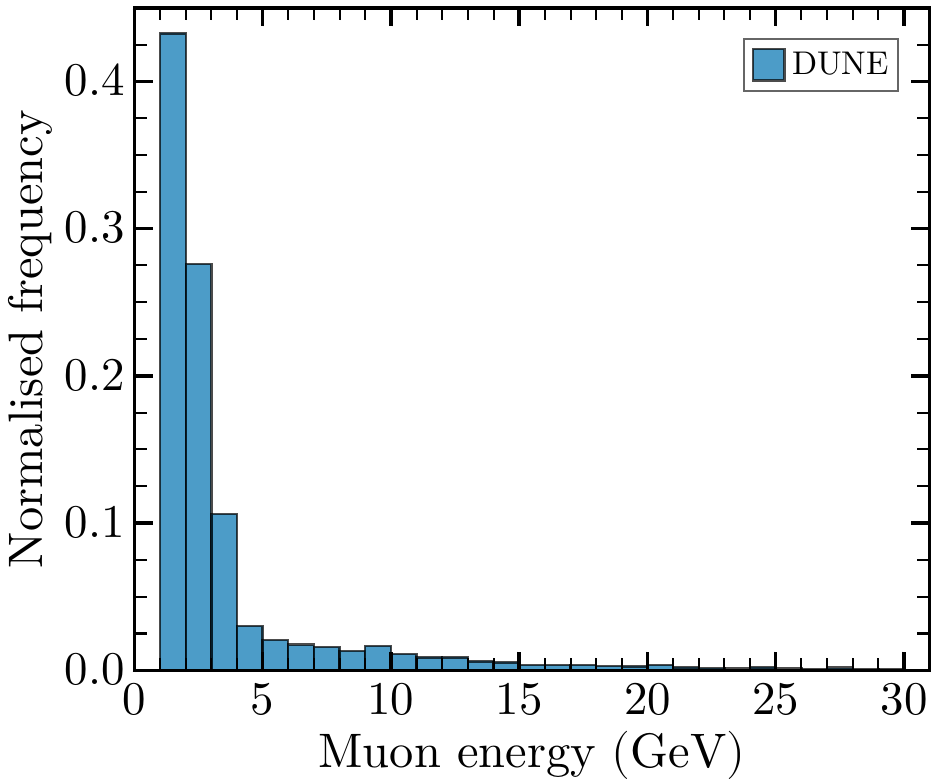}
  \end{subfigure}
  \hfill
  \begin{subfigure}{0.49\textwidth}
    \includegraphics[width=\textwidth]{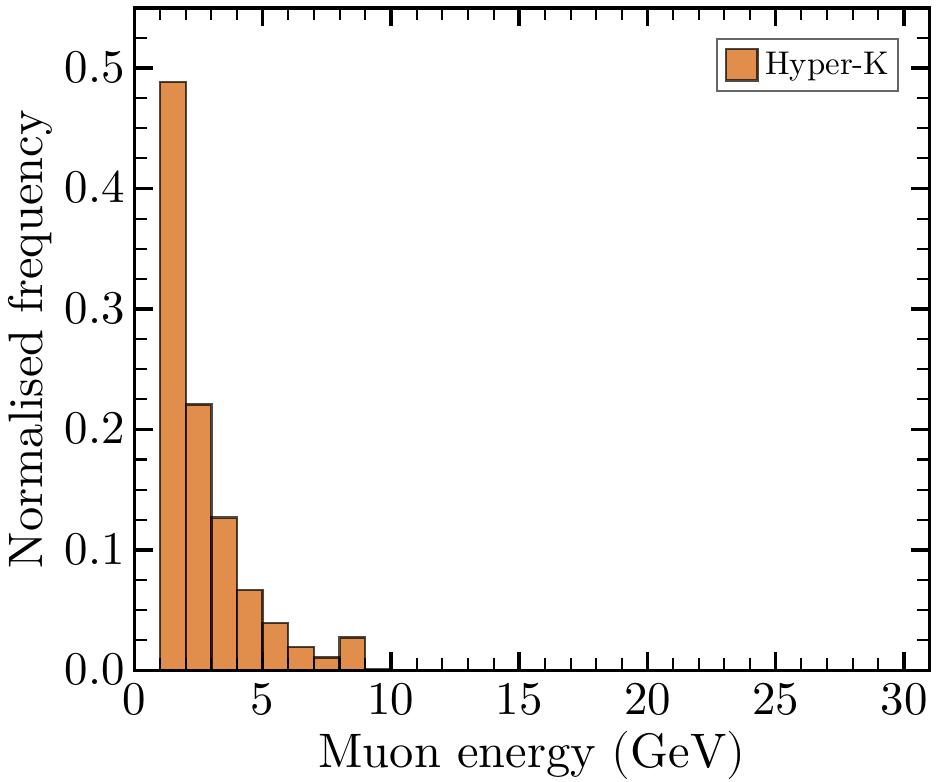}
  \end{subfigure}
  \caption{Left: Energy distribution of muons created by atmospheric neutrinos for DUNE. The right panel shows the analogous histogram for Hyper-Kamiokande.}
  \label{fig:BGs}
\end{figure}
A dominant background for the signal process, namely a high-energy antiproton created in the CR reaction, is muons created by atmospheric neutrinos. To determine the energy scale of these muons, we use atmospheric neutrinos fluxes from \cite{Honda:2015fha} and simulate the charged current events using the GiBUU generator \cite{Buss:2011mx}. We average over the azimuthal angle for the fluxes and consider only the muon neutrino component of the atmospheric flux. In Fig.~\ref{fig:BGs}, we present the normalised histograms of the energy distribution of the outgoing muons created from atmospheric neutrino interactions in the DUNE (left) and Hyper-Kamiokande (right) detector medium. We find that the muon energies produced at DUNE are higher than those at Hyper-Kamiokande. In addition to these muons produced from atmospheric neutrinos, cosmic muons could arrive at the detector. As described in \cite{Hyper-Kamiokande:2018ofw}, the average energy of cosmic muons in Hyper-K is around $\sim 200~\mathrm{GeV}$. These muons produce spallation background, but for energies above $26~\mathrm{MeV}$ no spallation background is expected, since the number of events decreases exponentially at higher energies. Still, hadronic interactions of high-energy muons interacting inside the detector could be a concern. However, that background can be rejected following the procedure described therein and considered negligible for our physics study case. To complement this, we have performed a simulation using CERN's FLUKA code \cite{fluka2025,Ahdida:2022gjl,Battistoni:2015epi} with the help of Flair's interface \cite{Donadon:2024omp} to study the passage of muons with an energy of $30~\mathrm{GeV}$ through the detector (both DUNE and Hyper-K). This beam of muons traverses a cylinder of 40 m of liquid argon for DUNE and a 73 m long water cylinder for Hyper-K. We find that muons traverse the whole detector, leaving a track that can be easily discarded, compared with our antiproton, which would be stopped on the inside of the detector.

\subsection{Expanding the detector with the Earth's crust}
\label{sec:earth-crust-expansion}
Since the produced antiproton has such high energies, we propose including part of the Earth's crust surrounding the detector as part of the target. The process should still be background-free if the antiproton reaches the detector with sufficient energy. For the DUNE Far Detector, we assume a detector mass of $m_{\mathrm{det}} = 40$ kilotons of liquid argon, while for Hyper-Kamiokande, a fiducial volume of $m_{\mathrm{det}} = 187$ kilotons of pure water is considered. From these volumes, the total number of electron targets $N_{\mathrm{target}}$ is computed as
\begin{equation}\label{eq:number-of-targets}
  N_{\mathrm{target}} = \dfrac{m_{\mathrm{det}} N_{\mathrm{A}}}{M_{\mathrm{target}}} N_e\,,
\end{equation}
where $N_{\mathrm{A}}$ is Avogadro's constant, $M_{\mathrm{target}}$ is the molar mass of the detector's material, and $N_e$ is the number of electrons in a molecule of the detector's material. Therefore, the number of targets in each detector is
\begin{align}
  N_{\mathrm{target}} \mathrm{~in~DUNE} &\approx 1.09\times10^{34}\,, \nonumber \\
  N_{\mathrm{target}} \mathrm{~in~Hyper\text{-}K} &\approx 6.25\times10^{34}\,. \nonumber
\end{align}
To enhance the number of signal events, we expand the detectors with an outer shell of rock (the Earth's crust) with a thickness of 1 meter, assuming that high-energy monopoles can scatter within this volume to produce energetic antiprotons. The Earth's crust is approximated to be made only of silicon (Si) with a density of $\rho = 2.7~\mathrm{g / cm^3}$. Knowing the density and volume of each shell, we can calculate its mass using that $m_{\mathrm{det}} = \rho V$ and substituting that into Eq.~\eqref{eq:number-of-targets}. For DUNE, each cavity\footnote{Assuming that the detector is not surrounded by rock but is located within a cave with some air.} has approximate dimensions of $72 \times 24 \times 21 \, \mathrm{m}^3$. Expanding the interaction region with a 1-meter thick parallelepiped shell\footnote{DUNE comprises four detectors. We multiply the volume of one cavity by four accordingly for the shell.} contributes approximately $2.58 \times 10^{34}$ additional targets within the crust. In the case of Hyper-Kamiokande, modelled as a cylinder with a height of 73 meters and a diameter of 69 meters, adding a cylindrical shell of 1-meter thickness yields an extra $1.62 \times 10^{34}$ targets. The greater number of additional targets in DUNE is attributed to the larger total volume of the four shells (approximately $3.2 \times 10^4 \, \mathrm{m}^3$) compared to the cylindrical shell of Hyper-K (approximately $2.0 \times 10^4 \, \mathrm{m}^3$). In summary, the total number of targets with the enhancement due to the additional volume provided by the shells is
\begin{align}
  N_{\mathrm{target}} \mathrm{~in~DUNE} &\approx 3.67\times10^{34}, \nonumber \\
  N_{\mathrm{target}} \mathrm{~in~Hyper}\text{-}\mathrm{K} &\approx 7.87\times10^{34}. \nonumber
\end{align}
Even though the enhancement is minor for Hyper-K, it is beneficial for DUNE, since it helps to approximately triplicate the total number of events when adding both contributions.

To justify this 1-meter shell expansion, we simulate a beam of 861.4 GeV antiprotons going through a 1-meter-long silicon cylinder. The beam then traverses two meters of air and collides with a liquid argon target. The simulation is performed again using CERN's FLUKA code and Flair's interface. This simulation aims to check that antiprotons produced within this silicon shell can escape and that they will arrive at the argon detector after travelling through some air, recreating the condition of DUNE's detector. In the case of Hyper-K, the shell is next to the detector, so if the simulation for DUNE results in the beam arriving at the liquid argon, the same will happen for Hyper-K. We find that 0.41 antiprotons/primary arrive at the detector with an energy of at least 30 GeV. We apply a lower energy cut of 30 GeV as the primary background from atmospheric neutrinos (see Fig.~\ref{fig:BGs}) has muons of maximal energy $\sim 20$ GeV. Even though not every antiproton arrives at the detector, additional charged hadrons are produced as secondary particles, carrying at least 30 GeV of energy. The total number of charged hadrons arriving to the detector (including the initial antiprotons) is 2.02 charged hadrons/primary. As we observe in Fig.~\ref{fig:BGs}, the number of background muons coming from atmospheric neutrinos with an energy of 30 GeV or higher is negligible. Therefore, our assumption of a 1-meter shell made of Si to increase the total number of events is reasonable. We proceed with a background-free analysis, since the monopole signal carries enough energy to be distinguished from that of a background one.
\subsection{Number of events and exclusion limits estimation}
\label{sec:number-of-events}
The number of signal events is given by
\begin{equation}
  N = N_{\mathrm{target}} \, t_{\mathrm{run}} \Phi \int_{t^-}^{t^+} \left(\dfrac{\mathrm{d}\sigma}{\mathrm{d}t}\right)\, \mathrm{d}t\,,
\end{equation}
where $N_{\mathrm{target}}$ is the number of targets (inside the detector plus the shell expansion as explained in \secref{sec:earth-crust-expansion}), $t_{\mathrm{run}}$ is the total exposure time and $\Phi$ is the monopole flux. This is a simplified analysis in which we assume a perfect detector, that is, one with an efficiency of $100\%$, and without accounting for any possible smearing effects, among others. To evaluate the 90\% C.L. sensitivity with a background-free set-up based only on the number of interaction events, we require $N > 2.3$ during the whole exposure time. This value of 2.3 comes from assuming Poisson statistics to determine the number of signal events required for a $90\%$ C.L. sensitivity in the absence of background.
\section{Results and discussion}
\label{sec:results-and-discussion}
In this section, we present the results of our analyses, starting with obtaining exclusion regions for the antiproton production and then estimating the flux of monopoles that would produce proton catalysis with a lifetime greater than the projected sensitivities of future experiments.

\subsection{Antiproton production}
\label{sec:results-antiproton}
\begin{figure}
\centering
\begin{subfigure}{0.49\textwidth}
  \includegraphics[width=\textwidth]{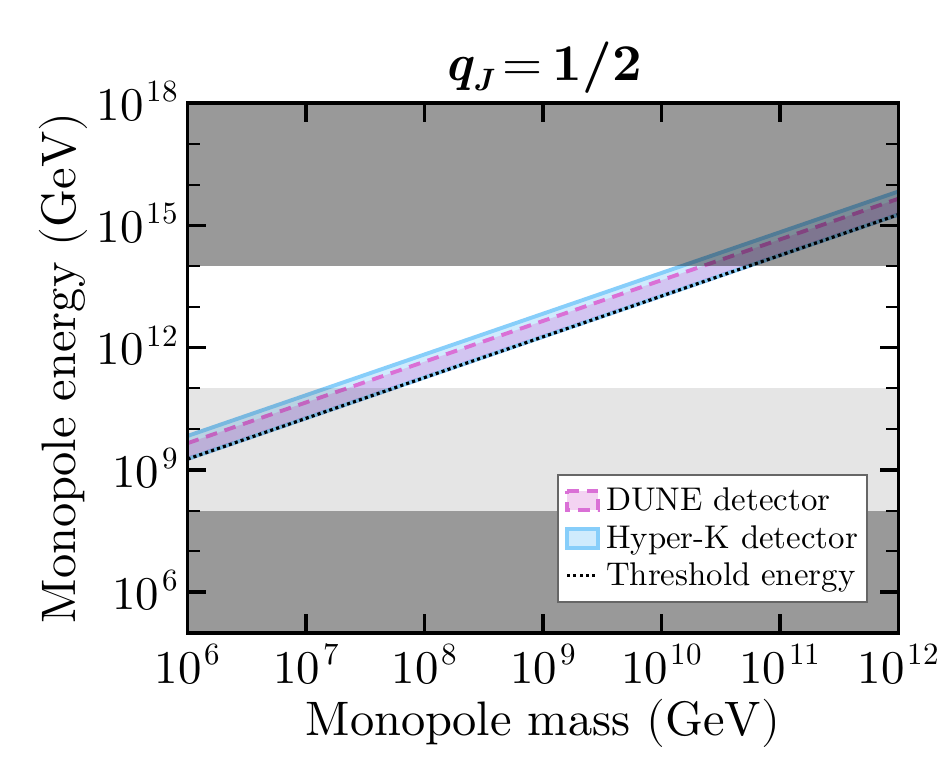}
\end{subfigure}
\hfill
\begin{subfigure}{0.49\textwidth}
  \includegraphics[width=\textwidth]{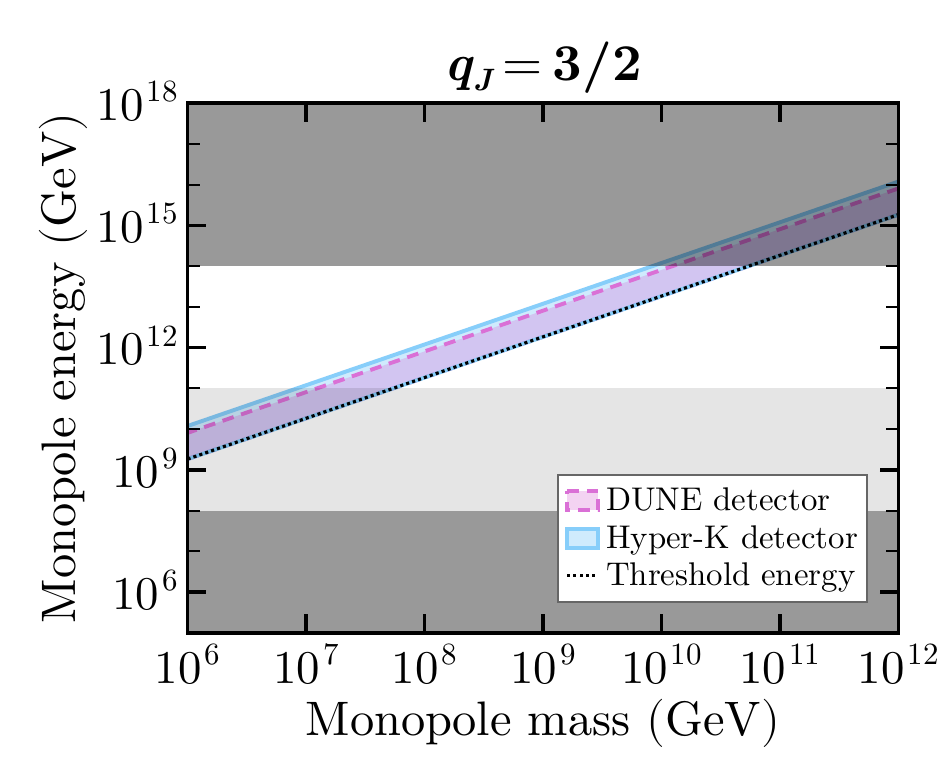}
\end{subfigure}

\begin{subfigure}{0.49\textwidth}
  \includegraphics[width=\textwidth]{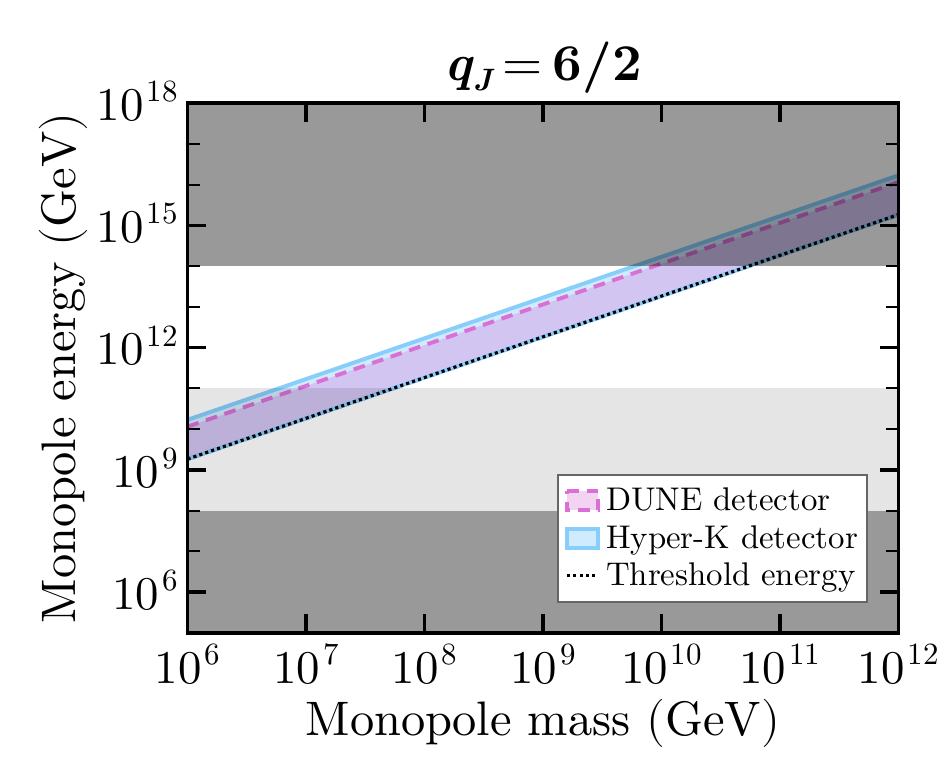}
\end{subfigure}
\hfill
\begin{subfigure}{0.49\textwidth}
  \includegraphics[width=\textwidth]{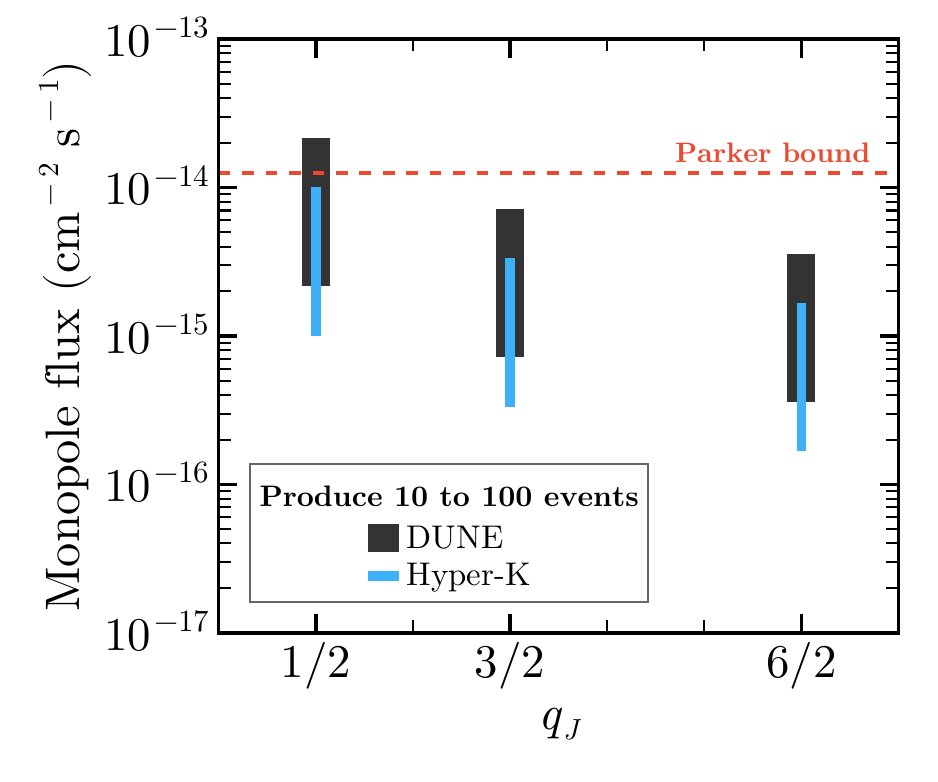}
\end{subfigure}
\caption{Top row and bottom-left panels: 90\% C.L. (2 d.o.f.) exclusion regions in the monopole energy-monopole mass plane for different values of $q_J$. Coloured bands depict the allowed values for the DUNE and Hyper-K detectors. Values below the threshold energy (dotted black line) are kinematically forbidden. The analysis is valid for the non-shaded region. The shaded dark grey region corresponds to monopoles that would never reach the detector (energies $< 10^8$ GeV) or to monopoles with an energy above the maximum energy that can be gained from extra-galactic fields (energies $> 10^{14}$ GeV). The shaded light grey region (energies between $10^8$ GeV and $10^{11}$ GeV) shows the region where only downward- and horizontal-going monopoles would make it to the detector.
Bottom-right panel: each bar represents the monopole flux that can produce between 10 and 100 events for the given $q_J$ and a monopole with $E_M^{\mathrm{lab}} = 1.89 \times 10^{12} ~\mathrm{GeV}$ and $m_M = 8.41 \times 10^8~\mathrm{GeV}$. The lower end of the bar is the 10 events limit and the upper end of it is the 100 events one. A dashed red line represents the Parker bound.} 
\label{fig:exclusion-limits}
\end{figure}

We start our analysis by fixing the monopole flux as $\Phi = 4\pi \times 10^{-16}~\mathrm{cm^{-2}~s^{-1}}$, one order of magnitude below the Parker bound. In addition, we assume a total experimental run time to be $15$ years. Next, we probe the parameter space for different monopole masses and energies (at the detector in the laboratory frame) and obtain the 90\% C.L. exclusion regions for different values of $q_J$. In the upper row and bottom-left panels of Fig.~\ref{fig:exclusion-limits} we represent the allowed values of energies and masses as coloured bands for the DUNE (dashed magenta bands) and Hyper-K (solid blue bands) detectors. In addition, the threshold energy of the process is marked as a dotted black line. As explained in \secref{sec:kin-synth}, monopoles with energies and masses falling below this line will not produce an antiproton if scattered off an electron in the detector, since they do not carry enough energy. The shaded dark grey region corresponds to monopoles that would never reach the detector, since they carry energies smaller than $10^8$~GeV, as explained in \secref{sec:outside-detector}. The upper shaded dark grey region represents monopoles with an energy greater than $10^{14}$~GeV, which is above the maximum energy that can be gained from extra-galactic fields. The analysis is not valid for monopoles with energies that fall inside these dark grey regions, since they will not arrive to the detector. The shaded light grey region, with energies between $10^8$~GeV and $10^{11}$~GeV, shows the region where only downward- and horizontal-going monopoles would make it to the detector. Inside the non-shaded region, the analysis is completely valid for every zenith angle that the incoming monopole may have.

By observing the coloured bands of the allowed values, we see that they are similar to the darkest regions of the top row of Fig.~\ref{fig:cs-and-momenta-comparison}, in which the highest cross-sections fall within a parallel band to the threshold energy next to it. This is a consequence of the fact that higher cross-sections imply a greater number of events in our simple analysis. When one increases the value of $q_J$, the cross-section increases and therefore the band becomes wider. Finally, let us note that the fact that Hyper-K band is wider is because the detector plus shell region of it contains a greater number of targets than that of DUNE.

We also want to explore how different monopole fluxes change the amount of scatterings produced. If we take a monopole mass of $m_M = 8.41 \times 10^8~\mathrm{GeV}$ and an energy of $E_M^{\mathrm{lab}} = 1.89 \times 10^{12} ~\mathrm{GeV}$, these produce the maximal value of the cross-section between $\sigma \sim 10^{-28}~\mathrm{cm^2}$ and $\sigma \sim 10^{-27}~\mathrm{cm^2}$, depending on the value of $q_J$. Given these values, we compute the flux needed to produce at least to 100 events. In the bottom-right panel of Fig.~\ref{fig:exclusion-limits} we plot such flux range for different values of $q_J$ for the DUNE and Hyper-K detectors. The lower end of the bar represents the flux needed to produce 10 events, whereas the upper end represents the flux for 100 events. The Parker bound is marked with a dashed red line. Increasing the value of $q_J$ increases the cross-section and thus, a smaller flux can be used to produce the same amount of events. In the case of DUNE and $q_J = 1/2$, it is impossible to produce 100 events with a flux smaller than the Parker bound.

\subsection{Proton decay catalysis}
\label{sec:results-proton-decay}
As shown in Eq.~\eqref{eq:ptoe}, a non-relativistic monopole can catalyse proton decay. The most common channel produces a $\pi^0$ in the final state, i.e., $M + p \rightarrow M + e^ + \pi^0$. The final state could also include $\pi^+ \pi^-$ pairs or even kaons and muons if other families are included. Nonetheless, we can recast the sensitivities of upcoming detectors to this baryon-number violating process.

The differential cross-section that describes this process for slow monopoles ($\beta < 0.2$) is given by Eq.~\eqref{eq:sig_pdec}. For $q_J = 1/2$, this cross-section is independent of the angle of scattering and, assuming a velocity $\beta = 10^{-3}$, we find that its value is
\begin{equation}
    \sigma_{Mp\rightarrow M e^+ \pi^0} \approx 1.4\times 10^{-21}~\mathrm{cm^2}\, .
\end{equation}
The proton lifetime sensitivity for the channel $p \rightarrow e^+ + \pi^0$ at $90\%$ C.L. after a run time of ten years is $\tau = 7.8\times 10^{34}$~yr for Hyper-K \cite{Hyper-Kamiokande:2018ofw} and $\tau \approx 1.7 \times 10^{34}$~yr for DUNE \cite{DUNE:2015lol}. Assuming that either Hyper-K or DUNE \textit{does not detect} proton decay, that is, that the proton lifetime is not shorter than the projected sensitivities, we can constrain the flux of the non-relativistic monopoles, $\Phi$, to be
\begin{equation}\label{eq:non-relativistic-bounds}
    \Phi = \dfrac{1}{4\pi}\dfrac{1}{\sigma_{Mp\rightarrow M e^+ \pi^0} \, \tau} \approx \begin{cases}
        2.3 \times 10^{-23}~\mathrm{cm^{-2}\,s^{-1}\,sr^{-1}}\quad \mathrm{for~Hyper\text{-}K}, \\
        1.1 \times 10^{-22}~\mathrm{cm^{-2}\,s^{-1}\,sr^{-1}}\quad \mathrm{for~DUNE},
    \end{cases}
\end{equation}
which is eight (seven) orders of magnitude smaller than the Parker bound for Hyper-K (DUNE). Super-Kamiokande set a lower limit on the proton lifetime at $\tau / B(p \rightarrow e^+ \pi^0) > 1.6 \times 10^{34}$~years at a $90\%$ C.L. \cite{Super-Kamiokande:2016exg}, which would give a flux one order of magnitude bigger than that of the Hyper-K one (around $1.1 \times 10^{-23}~\mathrm{cm^{-2}\,s^{-1}\,sr^{-1}}$). This is because the higher the proton lifetime is, the lower the monopole flux must be in order to not produce proton catalysis. These bounds of Eq.~\eqref{eq:non-relativistic-bounds} for non-relativistic monopoles that induce proton decay are close to others studied previously such as \cite{Super-Kamiokande:2012tld}.

It is interesting to consider that while the signature for a non-relativistic monopole inducing proton catalysis is the same as GUT gauge boson mediated proton decay, the kinematics would be very different. For instance, in a typical non-supersymmetric GUT model, the proton decays at rest and the outgoing states, $\pi^0$ and $e^+$, would be produced back-to-back. However, in the monopole-induced proton decay, since the final state is three-body and the monopole is moving, albeit slowly, the outgoing states would not be back-to-back.

\section{Summary}
\label{sec:conclusion}
In this work, we have explored the potential of future neutrino experiments, DUNE and Hyper-Kamiokande, to probe magnetic monopoles via Callan-Rubakov (CR) processes. We considered both relativistic and non-relativistic monopoles and analysed two key signatures: high-energy antiproton production and proton decay catalysis. For relativistic monopoles, we studied the CR process leading to the production of antiprotons with energies around 900 GeV and evaluated the relevant cross-sections. By including detector surrounding shell regions in our analysis, we showed that the effective detection volume is enhanced up to $238\%$ for DUNE and $25\%$ for Hyper-Kamiokande. This enhancement allows for meaningful sensitivity to monopole fluxes even below the Parker bound, one order of magnitude smaller with $10^{-16}~\mathrm{cm^{-2}\,s^{-1}\,sr^{-1}}$. We presented regions in the monopole energy–mass plane for different CR charges $q_J$, and demonstrated that Hyper-Kamiokande provides a slightly broader sensitivity range compared to DUNE due to its larger fiducial volume. Furthermore, we quantified the range of monopole fluxes required to produce observable events. We showed that future experiments can probe monopole fluxes as low as one order of magnitude below the Parker bound, depending on $q_J$. For non-relativistic monopoles, we recast the sensitivity of Hyper-Kamiokande and DUNE to proton decay via the CR process $M + p \rightarrow M + e^+ + \pi^0$. Assuming a monopole velocity of $\beta \sim 10^{-3}$ and a benchmark CR cross-section with $q_J = 1/2$, we showed that the non-observation of proton decay over a decade of running at Hyper-Kamiokande would constrain the monopole flux to $\Phi \lesssim 2.3 \times 10^{-23}~\mathrm{cm^{-2}\,s^{-1}\,sr^{-1}}$, several orders of magnitude below the Parker bound and comparable with existing limits from Super-Kamiokande. In the case of DUNE, the flux is constrained to $\Phi \lesssim 1.1 \times 10^{-22}~\mathrm{cm^{-2}\,s^{-1}\,sr^{-1}}$. We also emphasised that, although the final states of monopole-induced proton decay resemble those of GUT-mediated proton decay, their distinct kinematic configurations could provide a discriminating handle in the event of a positive detection.

\section*{Acknowledgments}
It is a pleasure to thank Josh Barrow and Yun Tse Tsai for a very helpful discussion on the uniqueness of the signal in DUNE. We would also like to thank Dimitrios K. Papoulias and Ivan Martinez-Soler for their insightful discussions on the atmospheric neutrino fluxes and the analysis.
We are grateful to Martin Bauer, Frank Krauss and Michael Spannowsky for useful comments on this project. JT would like to thank the Quantum Field Theory Centre at the University of Southern Denmark for their hospitality during the completion of this work.
PMC is supported by the Spanish grants PID2023-147306NB-I00 and CEX2023-001292-S (MCIU/AEI/10.13039/501100011033), as well as CIPROM/2021/054, CIACIF/2021/281 and CIBEFP/2023/65 (Generalitat Valenciana).

\begin{appendix}
\section{Monopole cross-section and its kinematics}
\label{sec:appendix-cross-sections}
The cross-section that describes the process $M\, e_R^- \rightarrow M\, \overline{p}_L$ of Eq.~\eqref{eq:sig_etwo} in the CoM frame is given by Eq.~\eqref{eq:sig_etop}. For convenience, let us write the expression again as
\begin{equation}
  \dfrac{\mathrm{d}\sigma}{\mathrm{d}\Omega} \,=\, \dfrac{1}{2} \dfrac{|\mathbf{p_{\overline{p}}^{\, cm}}|}{|\mathbf{p_{e}^{\, cm}}|} \dfrac{q_J^2}{|\mathbf{p_{e}^{\, cm}}|^{2}} \left[\sin\left(\dfrac{\theta^{\mathrm{cm}}}{2}\right)\right]^{4 |q_J|-2},
\end{equation}
where $|\mathbf{p_{e}^{\, cm}}|$ and $|\mathbf{p_{\overline{p}}^{\, cm}}|$ denote the centre-of-mass momenta of the electron and the antiproton, respectively, and $\theta^{\mathrm{cm}}$ is the polar angle, taken to be the angle between $\mathbf{p_{e}^{\, cm}}$ and $\mathbf{p_{\overline{p}}^{\, cm}}$. The values of the momenta are given by
\begin{equation}\label{eq:momentum-e-pbar-com}
  |\mathbf{p_{e}^{\, cm}}| = \dfrac{1}{2\sqrt{s}} \lambda^{1/2}(s,\, m_M^2, m_e^2), \qquad |\mathbf{p_{\overline{p}}^{\, cm}}| = \dfrac{1}{2\sqrt{s}} \lambda^{1/2}(s,\, m_M^2,\, m_p^2),
\end{equation}
where $s$ is given by Eq.~\eqref{eq:s-lab} and $\lambda(x,\, y,\, z)$ is the K\"all\'en lambda function, defined as
\begin{equation}\label{eq:kallen-lambda}
  \lambda(x, y, z) = (x - y - z)^2 - 4 y z.
\end{equation}

As explained in \secref{sec:p_synth}, $q_j$ can take several values. When $q_j = 1/2$, the cross-section is independent of the polar angle and the integration is straightforward. The solid angle contributes with a $4\pi$. On the other hand, when $q_j \geq 3/2$, there is a contribution coming from $\cos(\theta^{\mathrm{cm}})$. It is convenient to rewrite the cross-section using the trigonometric identity
\begin{equation}
  \left[\sin\left(\dfrac{\theta^{\mathrm{cm}}}{2}\right)\right]^{4|q_j|-2} = \left[\dfrac{1-\cos(\theta^{\mathrm{cm}})}{2}\right]^{2|q_j|-1},
\end{equation}
which holds for every possible value of $|q_j|$ since it is always\footnote{For a general $|q_j|$, one needs to account for the sign of $\sin(\theta^{\mathrm{cm}}/2)$. In our case, however, the exponent is always even.} $q_J \in \mathbb{Z} / 2$.

Next, we integrate out the azimuthal component of the solid angle from the cross-section, obtaining a $2\pi$ factor, \textit{i.e.},
\begin{equation}
  \dfrac{\mathrm{d}\sigma}{\mathrm{d}\cos(\theta)} = 2\pi \dfrac{\mathrm{d}\sigma}{\mathrm{d}\Omega},
\end{equation}
and perform a change of variables to go from $\cos(\theta^{\mathrm{cm}})$ to the Mandelstam variable $t$. The function to be integrated will be
\begin{equation}
  \dfrac{\mathrm{d}\sigma}{\mathrm{d}t} = \dfrac{\pi}{|\vec{p}_{e}^{\,\rm cm}| |\vec{p}_{\overline{p}}^{\,\rm cm}|} \dfrac{\mathrm{d}\sigma}{\mathrm{d}\Omega},
\end{equation}
and the variable $t$ is constrained by $t^- \leq t \leq t^+$. The limits are given by
\begin{equation}
\begin{aligned}
  t^\pm = 2m_M^2 - \dfrac{1}{2s} \Big[&(s + m_M^2 - m_e^2) (s + m_M^2 - m_p^2) \\[4pt]
  &\mp \lambda^{1/2}(s,\, m_M^2,\, m_e^2) \lambda^{1/2}(s,\, m_M^2,\, m_p^2)\Big].
\end{aligned}
\end{equation}
The final ingredient is to express $\cos(\theta^{\mathrm{cm}})$ in terms of the Mandelstam variables. The value of $s$ is fixed by Eq.~\eqref{eq:s-lab} and the value of $u$ can be expressed in terms of $t$ as $u = 2m_M^2 + m_e^2 + m_p^2 - s - t$. Finally, by working out the kinematics one obtains that
\begin{equation}
  \cos(\theta^{\mathrm{cm}}) = \dfrac{s(t - u) + (m_M^2 - m_e^2)(m_M^2 - m_p^2)}{\lambda^{1/2}(s,\, m_M^2,\, m_e^2)\, \lambda^{1/2}(s,\, m_M^2,\, m_p^2)}.
\end{equation}

For completeness, knowing how to transition from the lab to the centre-of-mass frame in which the collision will occur is needed. To do so, the system must be boosted by $\beta_{\mathrm{cm}}$, given by
\begin{equation}
  \beta_{\rm cm} = \dfrac{|\mathbf{p_{M}^{\, lab}}|}{E^{\mathrm{lab}}_{M} + m_e},
\end{equation}
where
\begin{equation}
  |\mathbf{p_{M}^{\, lab}}| = \dfrac{1}{2 m_e} \sqrt{\lambda(s,\, m_M^2,\, m_e^2)}
\end{equation}
is the initial three-momentum of the monopole in the laboratory frame. In order to plot the bottom row of Fig.~\ref{fig:cs-and-momenta-comparison}, we need to know the value of the antiproton momentum in the laboratory frame. Given that its energy in the centre-of-mass frame is
\begin{equation}
  E_{\overline{p}}^{\mathrm{cm}} = \dfrac{s + m_p^2 - m_M^2}{2\sqrt{s}},
\end{equation}
and its momentum is given by Eq.~\eqref{eq:momentum-e-pbar-com}, we can boost the system back to the laboratory frame using $\beta_{\mathrm{cm}}$ to obtain $\mathbf{p_{\overline{p}}^{lab}}$.

\section{Mean free path of monopoles}
\label{app:MFP}
We want to estimate the mean free path $\lambda$ of the monopole that scatters when traveling through the Earth following the process of Eq.~\eqref{eq:sig_etop}. The mean free path is calculated as
\begin{equation}
    \lambda = \dfrac{1}{\sigma\, n_{\mathrm{atoms}}},
\end{equation}
where $n_{\mathrm{atoms}}$ is the number density of atoms in the mantle and $\sigma$ is the total cross-section. To make a very rough estimate, let us take the density of the Earth's crust to be $\rho = 4~\mathrm{g / cm^3}$. The monopoles would be traveling isotropically through the more dense lower mantle and core, but for now, we are going to assume the density to be approximately that of the upper mantle. Note that the crust (around 5 km deep) is less dense. The mantle is composed mostly of MgO and SiO${}_2$, with molar masses of approximately $60~\mathrm{g/mol}$ and $40~\mathrm{g/mol}$, respectively. For simplicity, we can take an average molar mass of $M = 50~\mathrm{g/mol}$. Hence, then number density of atoms can be computed as
\begin{equation}
    n_{\text{atoms}} = \frac{\rho N_A}{M} = \frac{4 \, \text{g/cm}^{3} \cdot 6.02 \times 10^{23} \, \text{mol}^{-1}}{50 \, \text{g/mol}} \sim 4.8 \times 10^{22} \, \text{cm}^{-3},
\end{equation}
where $N_A$ is Avogadro's constant.

Next, we can estimate the number of electrons in SiO${}_2$. Taking that each molecule of SiO${}_2$ contains 30 electrons, the number density of electrons is
\begin{equation}
    n_{\mathrm{electrons}} \sim 1.4 \times 10^{24}~\mathrm{cm^{-3}}\, .
\end{equation}
For a given cross-section, we can estimate the mean free path of the monopole. Taking, for example, $\sigma \sim 10^{-28}~\mathrm{cm^2}$, one finds the mean free path for a typical cross-section to be
\begin{equation}
\lambda \sim \frac{1}{10^{-28} \, \text{cm}^{2} \cdot 1.4 \times 10^{24} \, \text{cm}^{-3}} \sim 70~\mathrm{m}\, .
\end{equation}
\end{appendix}

\bibliographystyle{JHEP}
\bibliography{ref}

\providecommand{\href}[2]{#2}\begingroup\raggedright\begin{thebibliography}{10}

\bibitem{Dirac:1931kp}
P.A.M.~Dirac, {{Quantised singularities in the electromagnetic field,}}, \href{https://doi.org/10.1098/rspa.1931.0130}{{Proc. Roy. Soc. Lond. A} {\bfseries 133} (1931) 60}.

\bibitem{tHooft:1974kcl}
G.~'t~Hooft, {{Magnetic Monopoles in Unified Gauge Theories}}, \href{https://doi.org/10.1016/0550-3213(74)90486-6}{{Nucl. Phys. B} {\bfseries 79} (1974) 276}.

\bibitem{Polyakov:1974ek}
A.M.~Polyakov, {{Particle Spectrum in Quantum Field Theory}}, {{JETP Lett.} {\bfseries 20} (1974) 194}.

\bibitem{Zeldovich:1978wj}
Y.B.~Zeldovich and M.Y.~Khlopov, {{On the Concentration of Relic Magnetic Monopoles in the Universe}}, \href{https://doi.org/10.1016/0370-2693(78)90232-0}{{Phys. Lett. B} {\bfseries 79} (1978) 239}.

\bibitem{Preskill:1979zi}
J.~Preskill, {{Cosmological Production of Superheavy Magnetic Monopoles}}, \href{https://doi.org/10.1103/PhysRevLett.43.1365}{{Phys. Rev. Lett.} {\bfseries 43} (1979) 1365}.

\bibitem{Guth:1980zm}
A.H.~Guth, {{The Inflationary Universe: A Possible Solution to the Horizon and Flatness Problems}}, \href{https://doi.org/10.1103/PhysRevD.23.347}{{Phys. Rev. D} {\bfseries 23} (1981) 347}.

\bibitem{MoEDAL:2019ort}
{\scshape MoEDAL} collaboration, {{Magnetic Monopole Search with the Full MoEDAL Trapping Detector in 13 TeV pp Collisions Interpreted in Photon-Fusion and Drell-Yan Production}}, \href{https://doi.org/10.1103/PhysRevLett.123.021802}{{Phys. Rev. Lett.} {\bfseries 123} (2019) 021802} [\href{https://arxiv.org/abs/1903.08491}{{\ttfamily 1903.08491}}].

\bibitem{Gould:2024zed}
O.~Gould, I.~Ostrovskiy and A.~Upreti, {{The next frontiers for magnetic monopole searches}},  \href{https://arxiv.org/abs/2409.04552}{{\ttfamily 2409.04552}}.

\bibitem{IceCube:2021eye}
{\scshape IceCube} collaboration, {{Search for Relativistic Magnetic Monopoles with Eight Years of IceCube Data}}, \href{https://doi.org/10.1103/PhysRevLett.128.051101}{{Phys. Rev. Lett.} {\bfseries 128} (2022) 051101} [\href{https://arxiv.org/abs/2109.13719}{{\ttfamily 2109.13719}}].

\bibitem{NOvA:2020qpg}
{\scshape NOvA} collaboration, {{Search for slow magnetic monopoles with the NOvA detector on the surface}}, \href{https://doi.org/10.1103/PhysRevD.103.012007}{{Phys. Rev. D} {\bfseries 103} (2021) 012007} [\href{https://arxiv.org/abs/2009.04867}{{\ttfamily 2009.04867}}].

\bibitem{IceCube:2014xnp}
{\scshape IceCube} collaboration, {{Search for non-relativistic Magnetic Monopoles with IceCube}}, \href{https://doi.org/10.1140/epjc/s10052-014-2938-8}{{Eur. Phys. J. C} {\bfseries 74} (2014) 2938} [\href{https://arxiv.org/abs/1402.3460}{{\ttfamily 1402.3460}}].

\bibitem{Super-Kamiokande:2012tld}
{\scshape Super-Kamiokande} collaboration, {{Search for GUT monopoles at Super\textendash{}Kamiokande}}, \href{https://doi.org/10.1016/j.astropartphys.2012.05.008}{{Astropart. Phys.} {\bfseries 36} (2012) 131} [\href{https://arxiv.org/abs/1203.0940}{{\ttfamily 1203.0940}}].

\bibitem{Zhang:2024mze}
C.~Zhang, S.-H.~Zhang, B.~Fu, J.-F.~Zhang and X.~Zhang, {{On the cosmological abundance of magnetic monopoles}}, \href{https://doi.org/10.1007/JHEP08(2024)220}{{JHEP} {\bfseries 08} (2024) 220} [\href{https://arxiv.org/abs/2404.04926}{{\ttfamily 2404.04926}}].

\bibitem{Parker:1970xv}
E.N.~Parker, {{The Origin of Magnetic Fields}}, \href{https://doi.org/10.1086/150442}{{Astrophys. J.} {\bfseries 160} (1970) 383}.

\bibitem{Hu:2022wcd}
H.~Hu, J.~Cheng, W.-L.~Guo and W.~Wang, {{Exploring neutrinos from proton decays catalyzed by GUT monopoles in the Sun}}, \href{https://doi.org/10.1088/1475-7516/2022/06/003}{{JCAP} {\bfseries 06} (2022) 003} [\href{https://arxiv.org/abs/2201.02386}{{\ttfamily 2201.02386}}].

\bibitem{DUNE:2020ypp}
{\scshape DUNE} collaboration, {{Deep Underground Neutrino Experiment (DUNE), Far Detector Technical Design Report, Volume II: DUNE Physics}},  \href{https://arxiv.org/abs/2002.03005}{{\ttfamily 2002.03005}}.

\bibitem{Hyper-Kamiokande:2018ofw}
{\scshape Hyper-Kamiokande} collaboration, {{Hyper-Kamiokande Design Report}},  \href{https://arxiv.org/abs/1805.04163}{{\ttfamily 1805.04163}}.

\bibitem{Rubakov:1981rg}
V.A.~Rubakov, {{Superheavy Magnetic Monopoles and Proton Decay}}, {{JETP Lett.} {\bfseries 33} (1981) 644}.

\bibitem{Rubakov:1982fp}
V.A.~Rubakov, {{Adler-Bell-Jackiw Anomaly and Fermion Number Breaking in the Presence of a Magnetic Monopole}}, \href{https://doi.org/10.1016/0550-3213(82)90034-7}{{Nucl. Phys. B} {\bfseries 203} (1982) 311}.

\bibitem{Callan:1982ah}
C.G.~Callan, Jr., {{Disappearing Dyons}}, \href{https://doi.org/10.1103/PhysRevD.25.2141}{{Phys. Rev. D} {\bfseries 25} (1982) 2141}.

\bibitem{Callan:1982au}
C.G.~Callan, Jr., {{Dyon-Fermion Dynamics}}, \href{https://doi.org/10.1103/PhysRevD.26.2058}{{Phys. Rev. D} {\bfseries 26} (1982) 2058}.

\bibitem{Rubakov:1988aq}
V.A.~Rubakov, {{Monopole Catalysis of Proton Decay}}, \href{https://doi.org/10.1088/0034-4885/51/2/002}{{Rept. Prog. Phys.} {\bfseries 51} (1988) 189}.

\bibitem{Khoze:2024hlb}
V.V.~Khoze, {{Monopoles and fermions in the Standard Model}}, \href{https://doi.org/10.1007/JHEP09(2024)146}{{JHEP} {\bfseries 09} (2024) 146} [\href{https://arxiv.org/abs/2405.18689}{{\ttfamily 2405.18689}}].

\bibitem{Kazama:1976fm}
Y.~Kazama, C.N.~Yang and A.S.~Goldhaber, {{Scattering of a Dirac Particle with Charge Ze by a Fixed Magnetic Monopole}}, \href{https://doi.org/10.1103/PhysRevD.15.2287}{{Phys. Rev. D} {\bfseries 15} (1977) 2287}.

\bibitem{Csaki:2020inw}
C.~Cs\'aki, S.~Hong, Y.~Shirman, O.~Telem, J.~Terning and M.~Waterbury, {{Scattering amplitudes for monopoles: pairwise little group and pairwise helicity}}, \href{https://doi.org/10.1007/JHEP08(2021)029}{{JHEP} {\bfseries 08} (2021) 029} [\href{https://arxiv.org/abs/2009.14213}{{\ttfamily 2009.14213}}].

\bibitem{Csaki:2021ozp}
C.~Cs\'aki, Y.~Shirman, O.~Telem and J.~Terning, {{Pairwise Multiparticle States and the Monopole Unitarity Puzzle}}, \href{https://doi.org/10.1103/PhysRevLett.129.181601}{{Phys. Rev. Lett.} {\bfseries 129} (2022) 181601} [\href{https://arxiv.org/abs/2109.01145}{{\ttfamily 2109.01145}}].

\bibitem{Khoze:2023kiu}
V.V.~Khoze, {{Scattering amplitudes of fermions on monopoles}}, \href{https://doi.org/10.1007/JHEP11(2023)214}{{JHEP} {\bfseries 11} (2023) 214} [\href{https://arxiv.org/abs/2308.09401}{{\ttfamily 2308.09401}}].

\bibitem{vanBeest:2023dbu}
M.~van Beest, P.~Boyle~Smith, D.~Delmastro, Z.~Komargodski and D.~Tong, {{Monopoles, scattering, and generalized symmetries}}, \href{https://doi.org/10.1007/JHEP03(2025)014}{{JHEP} {\bfseries 03} (2025) 014} [\href{https://arxiv.org/abs/2306.07318}{{\ttfamily 2306.07318}}].

\bibitem{vanBeest:2023mbs}
M.~van Beest, P.~Boyle~Smith, D.~Delmastro, R.~Mouland and D.~Tong, {{Fermion-monopole scattering in the Standard Model}}, \href{https://doi.org/10.1007/JHEP08(2024)004}{{JHEP} {\bfseries 08} (2024) 004} [\href{https://arxiv.org/abs/2312.17746}{{\ttfamily 2312.17746}}].

\bibitem{Wick:2000yc}
S.D.~Wick, T.W.~Kephart, T.J.~Weiler and P.L.~Biermann, {{Signatures for a cosmic flux of magnetic monopoles}}, \href{https://doi.org/10.1016/S0927-6505(02)00200-1}{{Astropart. Phys.} {\bfseries 18} (2003) 663} [\href{https://arxiv.org/abs/astro-ph/0001233}{{\ttfamily astro-ph/0001233}}].

\bibitem{Perri:2023ncd}
D.~Perri, K.~Bondarenko, M.~Doro and T.~Kobayashi, {{Monopole acceleration in intergalactic magnetic fields}}, \href{https://doi.org/10.1016/j.dark.2024.101704}{{Phys. Dark Univ.} {\bfseries 46} (2024) 101704} [\href{https://arxiv.org/abs/2401.00560}{{\ttfamily 2401.00560}}].

\bibitem{PhysRev.138.B248}
D.R.~Tompkins, {Total energy loss and \ifmmode \check{C}\else \v{C}\fi{}erenkov emission from monopoles}, \href{https://doi.org/10.1103/PhysRev.138.B248}{{Phys. Rev.} {\bfseries 138} (1965) B248}.

\bibitem{Derkaoui:1998uv}
J.~Derkaoui, G.~Giacomelli, T.~Lari, A.~Margiotta, M.~Ouchrif, L.~Patrizii et~al., {{Energy losses of magnetic monopoles and of dyons in the earth}}, \href{https://doi.org/10.1016/S0927-6505(98)00016-4}{{Astropart. Phys.} {\bfseries 9} (1998) 173}.

\bibitem{Honda:2015fha}
M.~Honda, M.~Sajjad~Athar, T.~Kajita, K.~Kasahara and S.~Midorikawa, {{Atmospheric neutrino flux calculation using the NRLMSISE-00 atmospheric model}}, \href{https://doi.org/10.1103/PhysRevD.92.023004}{{Phys. Rev. D} {\bfseries 92} (2015) 023004} [\href{https://arxiv.org/abs/1502.03916}{{\ttfamily 1502.03916}}].

\bibitem{Buss:2011mx}
O.~Buss, T.~Gaitanos, K.~Gallmeister, H.~van Hees, M.~Kaskulov, O.~Lalakulich et~al., {{Transport-theoretical Description of Nuclear Reactions}}, \href{https://doi.org/10.1016/j.physrep.2011.12.001}{{Phys. Rept.} {\bfseries 512} (2012) 1} [\href{https://arxiv.org/abs/1106.1344}{{\ttfamily 1106.1344}}].

\bibitem{fluka2025}
``{FLUKA CERN}.'' \url{https://fluka.cern}.

\bibitem{Ahdida:2022gjl}
C.~Ahdida et~al., {{New Capabilities of the FLUKA Multi-Purpose Code}}, \href{https://doi.org/10.3389/fphy.2021.788253}{{Front. in Phys.} {\bfseries 9} (2022) 788253}.

\bibitem{Battistoni:2015epi}
G.~Battistoni et~al., {{Overview of the FLUKA code}}, \href{https://doi.org/10.1016/j.anucene.2014.11.007}{{Annals Nucl. Energy} {\bfseries 82} (2015) 10}.

\bibitem{Donadon:2024omp}
A.~Donadon, G.~Hugo, C.~Theis and V.~Vlachoudis, {{FLAIR3 \textendash{} recasting simulation experiences with the Advanced Interface for FLUKA and other Monte Carlo codes}}, \href{https://doi.org/10.1051/epjconf/202430211005}{{EPJ Web Conf.} {\bfseries 302} (2024) 11005}.

\bibitem{DUNE:2015lol}
{\scshape DUNE} collaboration, {{Long-Baseline Neutrino Facility (LBNF) and Deep Underground Neutrino Experiment (DUNE)}: {Conceptual Design Report, Volume 2: The Physics Program for DUNE at LBNF}},  \href{https://arxiv.org/abs/1512.06148}{{\ttfamily 1512.06148}}.

\bibitem{Super-Kamiokande:2016exg}
{\scshape Super-Kamiokande} collaboration, {{Search for proton decay via $p \to e^+\pi^0$ and $p \to \mu^+\pi^0$ in 0.31 megaton\textperiodcentered{}years exposure of the Super-Kamiokande water Cherenkov detector}}, \href{https://doi.org/10.1103/PhysRevD.95.012004}{{Phys. Rev. D} {\bfseries 95} (2017) 012004} [\href{https://arxiv.org/abs/1610.03597}{{\ttfamily 1610.03597}}].

\end{thebibliography}\endgroup

\end{document}